\documentclass[%
 reprint,
  superscriptaddress,
 amsmath,amssymb,amsfonts,
 aps,
 prl,
 floatfix,
 longbibliography
]{revtex4-2}

\usepackage[utf8]{inputenc}

\usepackage[pdftex]{graphicx} \graphicspath{{}}
\usepackage{float} 
\usepackage{xcolor}
\usepackage{verbatim}

\usepackage[pdftex,colorlinks=true]{hyperref}
\hypersetup{
  colorlinks=true,
  linkcolor=blue,
  urlcolor=cyan,
}

\usepackage{bm}

\usepackage{mathtools}

\DeclarePairedDelimiterX\braket[2]{\langle}{\rangle}{#1 \delimsize\vert #2}
\DeclarePairedDelimiterX\expval[3]{\langle}{\rangle}{#1 \delimsize\vert #2  \delimsize\vert #3}
\DeclarePairedDelimiterX\proj[2]{\delimsize\vert#1\rangle}{\langle#2\delimsize\vert}{ }

\begin{document}

\title{Two-mode squeezing in Floquet engineered power-law interacting spin models}

\author{Arman Duha}
\affiliation{Department of Physics, Oklahoma State University, Stillwater, Oklahoma 74078, USA}

\author{Thomas Bilitewski}
\affiliation{Department of Physics, Oklahoma State University, Stillwater, Oklahoma 74078, USA}

\date{\today}

\begin{abstract}
We study the non-equilibrium dynamics of a quantum spin 1/2 XXZ model confined in a two-dimensional bi-layer system, with couplings mediated by inverse power-law interactions, falling off with distance r as $1/r^{\alpha}$, with spatio-temporal control of the spins enabled via local fields.
An initial state of spins with opposite magnetization in the two layers is dynamically unstable resulting in exponential generation of correlated pairs of excitations. We find that scalabale generation of entanglement in the form of two-mode squeezing between the layers can generically be achieved in powerlaw models. 
We further demonstrate that spatially-temporally engineered interactions allow to significantly increase the generated entanglement and in fact achieve Heisenberg limited scaling. 
This work is relevant to a wide variety of experimental atomic, molecular, and optical platforms, which realise powerlaw spin models, and demonstrates the advantage of spatio-temporal control to maximise the generation of metrologically useful entanglement, with potential applications in quantum-enhanced sensing.
\end{abstract}

\maketitle

{\it Introduction.---} %
Long-range interacting spin-models realized in quantum gases of polar molecules \cite{Baranov2012,Bohn2017,Moses2017}, magnetic atoms \cite{Chomaz2023}, Rydberg atoms \cite{Browaeys2020,Saffman2010} or trapped ions \cite{RevModPhys.93.025001} are emerging as promising platforms for quantum simulation \cite{Bloch2012,Gross2017,Daley2022,RevModPhys.93.025001}, computation \cite{Briegel2000,Morgado2021,Henriet2020} and quantum metrology \cite{RevMod_Metrology_2018}. These systems with the advent of quantum gas microscopes and tweezers \cite{Kaufman2021,Gross2021,anderegg2019optical,Christakis2023,holland2023demand,bao2023} now combine the presence of long-range interactions with the capability to control, manipulate and measure at the single-particle level in a site-resolved fashion.
In addition to spatial control and resolution the ability to design the interactions is required to realize fully programmable quantum simulators. A particularly powerful approach to the engineering of spin interactions is time-periodic driving or Floquet-engineering \cite{Oka2019,Weitenberg2021}, and control of the interactions via pulse-sequences \cite{Lukin_2020_Robust,PhysRevLett.131.220803}. Indeed, this temporal control enables the desired programmability of the interactions as demonstrated in cavities \cite{Periwal2021}, Rydberg atoms \cite{PhysRevA.108.053318,Scholl2022,Geier2021}, polar molecules \cite{Christakis2023,miller2024twoaxis} and trapped ions \cite{PRXQuantum.4.010334}.

One particular direction in the quest for quantum advantage enabled by these advances is the enhancement of sensitivity of measurements via entanglement in the form of spin squeezing \cite{Wineland1992,wineland1994}. While spin squeezing had been achieved for infinite range interactions and via QND measurements \cite{vasilakis2015generation,appel2009mesoscopic,schleier2010states,bohnet2014reduced,sewell2012magnetic,bao2020spin}, only recently have experiments demonstrated this for finite range interactions in trapped-ions and Rydberg arrays \cite{Eckner2023,Franke2023a,Bornet2023a,Hines2023}. The majority of these works has thus far focused on homogeneous globally collective initial states interacting via homogeneous interactions and global controls, not exploiting the advantage offered by fully controllable quantum platforms.

We go beyond this paradigm and demonstrate how spatio-temporal control, i.e. spatially dependent temporal control, applied to spatially patterned, structured initial states confer additional tangible benefits for entanglement generation. %
\begin{figure}
\includegraphics[width=\columnwidth]{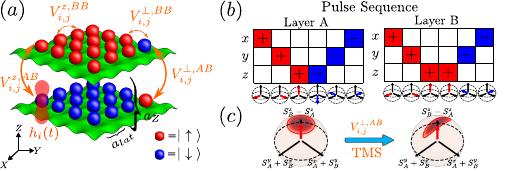}
\caption{(a) Illustration of spin 1/2 model confined in a bilayer interacting via powerlaw interactions with Ising $V^z$ and spin-exchange $V^{\perp}$, and intra-layer $(AA,BB)$ and inter-layer ($AB$) terms. Spins can be locally addressed via fields $h_{\boldsymbol{i}}(t)$ (b) Layer pulse-sequence in toggling frame (see text) to symmetrise intra-layer Ising, and cancel inter-layer Ising interactions. (c) Bloch sphere of mixed quadratures that are squeezed by the inter-plane spin-exchange interactions.\label{fig:fig1}}
\end{figure} 
Specifically, we demonstrate how to achieve Heisenberg limited scaling of two-mode squeezing via Floquet engineered spatially-anisotropic interactions in bilayers of power-law interacting XXZ spin models. Two-mode squeezing (TMS) \cite{Agarwal2013,Caves1985,Schumaker1985} is a process generating entanglement via the production of pairs of particles, a process underlying phenomenona from the fundamental in the EPR paradox \cite{Reid2009}, over Unruh thermal radiation \cite{Hu2019} and the Schwinger effect \cite{Hauke2013,Kasper2016} to the applied in parametric amplification in quantum optics \cite{Agarwal2013}. Two-mode squeezing is well established in photonic systems \cite{PhysRevLett.59.2555,Agarwal2013} and has been realized in thermal gases \cite{Julsgaard_2001,cerf2007quantum}, and in Bose-Einstein condensates of ultracold atoms interacting via contact interactions \cite{Gross2011,Luecke2011,bookjans2011strong,black2007spinor,zhao2014dynamics,qu2020probing,kim2021emission}, where it has been successfully used for EPR steering \cite{Lange2018,Fadel2018,Kunkel2018}. 
More recently it has been proposed as a mechanism to generate metrologically useful entanglement between spin ensembles in cavities via light-mediated interactions \cite{Sundar2023}, and in dipolar systems \cite{Bilitewski2023a,Bilitewski2023}. There long-range interactions naturally create entanglement in spatially separated ensembles, with the additional benefit of single-site single-particle control over these entangled states.

Here, we investigate the two-mode squeezing dynamics of spin 1/2 bilayers with powerlaw interactions, scaling as $r^{-\alpha}$ with the distance $r$. We find that finite-range interactions ($\alpha = 1, 2, 3$), achieve the same amount of squeezing as infinite-range ($\alpha = 0$) interactions for sufficiently widely separated layers. By Floquet engineering spatially anisotropic interactions adapted to the initial state and desired entanglement dynamics we improve the $1/\sqrt{N}$ scaling of the sensitivity in system size $N$ to the ultimate Heisenberg limit of $1/N$. %

Our work thus opens up new opportunities to exponentially generate metrologically useful entanglement in a variety of platforms including Rydberg atoms ($\alpha = 3,6$) \cite{Browaeys2020,Saffman2010}, polar molecules ($\alpha = 3$) \cite{Bohn2017,Moses2017,Baranov2012}, magnetic atoms ($\alpha = 3$) \cite{Chomaz2023}, and trapped ions ($1<\alpha <3$) \cite{RevModPhys.93.025001} wherin single-site \cite{Christakis2023}, layer selective control \cite{Tobias2022}, and Floquet-engineering \cite{miller2024twoaxis,Lukin_2020_Robust,PhysRevA.108.053318} have been demonstrated, and which can realise bilayer structures \cite{hawaldar2024bilayer,miller2024twoaxis,Meng2023,Tobias2022}.

{\it Model.---} %
We consider a two-dimensional bilayer of spins interacting via long-range interactions as shown in Fig. \ref{fig:fig1}(a). The two layers, denoted as $A$ and $B$, both have a square geometry with lattice spacing $a_{lat}$ and are separated by a tunable distance $a_Z$. The spins have two internal states and the dynamics is governed by a spin 1/2 XXZ Hamiltonian
\begin{align}\label{H1}
    \hat{H}=& 1/2\sum_{\boldsymbol{i}\neq \boldsymbol{j}} V_{\boldsymbol{i}\boldsymbol{j}} \Bigg[\frac{V_{\perp}}{2}(\hat{s}_{\boldsymbol{i}}^+ \hat{s}_{\boldsymbol{j}}^- +  \hat{s}_{\boldsymbol{i}}^- \hat{s}_{\boldsymbol{j}}^+ ) + V_z \hat{s}_{\boldsymbol{i}}^z \hat{s}_{\boldsymbol{j}}^z \Bigg] \nonumber  \\
    & + \sum_{\boldsymbol{i}} h_{\boldsymbol{i},x}(t) \hat{s}_{\boldsymbol{i}}^x +h_{\boldsymbol{i},y}(t) \hat{s}_{\boldsymbol{i}}^y  +h_{\boldsymbol{i},z}(t) \hat{s}_{\boldsymbol{i}}^z 
\end{align}    
where $\boldsymbol{i}$, $\boldsymbol{j}$ are three-dimensional indices ($i_X, i_Y, i_Z$), specifying the layer-index $i_Z$ and positions $i_X$, $i_Y$ in the layer of size $L \times L$ with $N=L^2$ spins in each layer. The spin operators $\hat{s}_{\boldsymbol{i}}^{\mu}=\hat{\sigma}_{\boldsymbol{i}}^{\mu}/2$, with the  Pauli matrices $\hat{\sigma}^{\mu}$, act on the spin at site $\boldsymbol{i}$. We consider power-law interactions of the form $V_{\boldsymbol{i}\boldsymbol{j}} = |\boldsymbol{r_{i}} - \boldsymbol{r_{j}}|^{-\alpha}$ and $V_{\perp}$ and $V_z$ describe the relative strengths of the spin-exchange and Ising interactions. 
While these interactions only depend on distance to appeal to a range of systems, spatial anisotropy, as natural for dipolar ($\alpha =3$) interactions and studied before  \cite{Bilitewski2023}, is not expected to change the results of this work. We include time and position dependent fields to control the spins in the second line of Eq.~\eqref{H1}.

{\it Engineering optimal two-mode squeezing.---}
%
Starting from oppositely polarized layers, $\Vec{S}_{\eta} = \sum_{\boldsymbol{i}\in \eta}\Vec{s}_{\boldsymbol{i}} = \pm N/2 \hat{z}$ the goal is to utilize spin-spin interactions to generate entangled pairs of collective excitations between the layers.

The initial state is an eigenstate of intra-layer Heisenberg ($\vec{s}_{\boldsymbol{i}} \cdot \vec{s}_{\boldsymbol{j}}$) interactions which will protect the collective layer-spin by energetically penalizing non-collective excitations. In contrast, the interlayer spin-exchange interactions will create pairs of flipped spins, and the inter-layer Ising interactions will energetically penalize the creation of spin flips \cite{Bilitewski2023a,Bilitewski2023}, eventually arresting pair production. Thus, Heisenberg intra-layer interactions are required to realize collective dynamics (for finite range interactions), whereas Ising interactions between layers are detrimental. This suggests two strategies: (i) utilising Heisenberg interactions and canceling the inter-layer Ising interactions to leading order by a static field, or (ii) fully removing only the Ising inter-layer interactions, while keeping Heisenberg intra-layer interactions.

The first strategy corresponds to fully-symmetric Heisenberg interactions ($V_z = V_{\perp}$) and a staggered layer-dependent z-field, i.e.
\begin{equation}
\hat{H}_h = 1/2 \sum_{\boldsymbol{i}\boldsymbol{j}} V_{\boldsymbol{i}\boldsymbol{j}} \, \vec{s}_{\boldsymbol{i}} \cdot \vec{s}_{\boldsymbol{j}} + h (\hat{S}^z_B - \hat{S}^z_A)
\end{equation}
 where the local $z$ field exactly cancels the energy cost due to the inter-layer Ising interactions of creating one collective spin-flip in each layer on top of the initial state, $h = 1/(2N) \sum_{\boldsymbol{i}\in A,\boldsymbol{j} \in B} V_{\boldsymbol{i}\boldsymbol{j}} = N V_{\mathrm{avg}}/2$.

The second strategy corresponds to engineering
\begin{equation}
\hat{H}_{V} = 1/2 \sum_{\eta} \sum_{\boldsymbol{i},\boldsymbol{j} \in \eta} V_{\boldsymbol{i}\boldsymbol{j}} \, \vec{s}_{\boldsymbol{i}} \cdot \vec{s}_{\boldsymbol{j}} +  \sum_{\mathclap{\boldsymbol{i} \in A, \boldsymbol{j} \in B}} \, V_{\boldsymbol{i}\boldsymbol{j}} \, (\hat{s}_{\boldsymbol{i}}^x \hat{s}_{\boldsymbol{j}}^x +\hat{s}_{\boldsymbol{i}}^y \hat{s}_{\boldsymbol{j}}^y ) \label{eq:Vz}
\end{equation}
where the first term describes intra-layer Heisenberg, and the second term inter-layer spin-exchange interactions. 

We achieve this by extending interaction design via global pulses \cite{Lukin_2020_Robust} to layer-dependent pulse sequences utilizing local control of the spins. %
Starting from Ising interactions, $\hat{s}_{\boldsymbol{i}}^z \hat{s}_{\boldsymbol{j}}^z$, applying a rotation from $z$ to direction $\alpha (\beta)$ to the first (second) spin transforms the interactions to $\hat{s}_{\boldsymbol{i}}^{\alpha} \hat{s}_{\boldsymbol{j}}^{\beta}$. The resulting interactions under a sequence of rotations can therefore be represented by the orientation of the $z$-operator during the $k$-th step of the sequence, i.e. in the toggling-frame representation \cite{Lukin_2020_Robust}. Using the 6 step layer-dependent sequence illustrated in Fig.~\ref{fig:fig1}(b), Ising interactions are successively transformed into $xx$, $yy$ and $zz$ interactions in steps 1 to 3, both within and between the layers. Crucially, in step 4 by applying different pulses to layers A and B we obtain positive intra-layer $zz$, but negative inter-layer $zz$ interactions, followed by $yy$ and $xx$ interactions in steps 5 and 6. Thus, this sequence fully symmetrises the intra-plane Ising to Heisenberg interactions, and only generates $xx+yy$ inter-layer interactions fromn the original interlayer Ising interactions. The average Hamiltonian then is Eq.~\ref{eq:Vz} with an overall prefactor of $V_z/3$, which is absorbed in $V_{\mathrm{avg}}$. 

In the following we will work directly with the target Hamiltonian, rather than the explicit pulse sequence. We have checked convergence of the dynamics under this multi-step Floquet protocol to the desired dynamics as a function of the Floquet period \cite{supplemental}, which shows good agreement if the total sequence takes about half of a nearest neighbour interaction time, $T \approx 0.6 V_{nn}/\hbar$, which is well within reach of experimental platforms \cite{Lukin_2020_Robust,miller2024twoaxis}.


For both scenarios the fully collective regime can be understood in terms of Holstein-Primakoff bosons \cite{Holstein1940} of the collective layer spins, which in first order are $S_A^z = -N/2+ \hat{a}^{\dagger} \hat{a}$, $S_A^+ =  \hat{a}$, $S_A^- = \hat{a}^{\dagger}$, $S_B^z = N/2 - \hat{b}^{\dagger} \hat{b}$, $S_B^- =  \hat{b}$ and $S_B^+ =  \hat{b}^{\dagger}$. In terms of these we obtain
$ H_{\mathrm{TMS}} = N V_{\mathrm{avg}}  (\hat{a}^{\dagger} \hat{b}^{\dagger} + \hat{a} \hat{b})/2$
which is the desired two-mode squeezing hamiltonian \cite{Agarwal2013,Schumaker1985,Caves1985}. The number of generated entangled excitations $n_{ex} = \hat{a}^\dagger \hat{a} + \hat{b}^{\dagger} \hat{b} = S_A^z - S_B^z+N$ within TMS is predicted to grow exponentially $n_{ex} = 2\sinh^2(N V_{\mathrm{avg}} t/(2\hbar))$ \cite{Agarwal2013,Schumaker1985,Caves1985}. The generated entanglement results in squeezing of mixed quadratures between the collective layer spins. The squeezed quadratures correspond to $S^x_A + S^y_B$ and $S^y_A-S^x_B$, while the anti-squeezed quadratures correspond to $S^x_A - S^y_B$ and $S^y_A + S^x_B$. These show exponentially decreasing or growing variances $\mathrm{Var}[\mathcal{O}^{\pm}] =N/2 \, e^{\pm N V_{\mathrm{avg}}t/\hbar}$. We illustrate these quadratures on the Bloch-sphere for the initial state, as well as the squeezed state in Fig.~\ref{fig:fig1}(c).

The main difference between these two approaches is that in the staggered field case the inter-layer Ising interactions were canceled by the h-field up to quadratic order, whereas in the Floquet-engineered case the Ising interactions were removed at the Hamiltonian level. Thus, in quadratic order both approaches are equivalent, however, we will show that the second approach provides dramatic improvements over the first.

\begin{figure}[t!]
\includegraphics[width=\columnwidth]{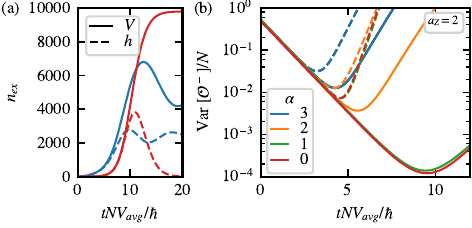}
\caption{Pair creation and two-mode squeezing via Floquet engineered spatially anisotropic interactions. (a) Number of created excitations $n_{ex} =S_A^z - S_B^z+N$ (b) Time evolution of variance of the squeezed quadrature. In both panels for the Floquet-engineered model (solid lines,$V$), and for the staggered field model (dashed, $h$), for a range of interaction powerlaw exponents $\alpha$ at fixed layer distance $a_Z=2$. 
\label{fig:fig2}}
\end{figure}

To substantiate the arguments based on the mapping to bosons in the collective limit we simulate the quantum many-body non-equilibrium dynamics of the spin model using the discrete truncated Wigner approximation (dTWA) \cite{Schachenmayer2015,Zhu_NewJournalofPhysics_21_2019} for up to $2N = 2\times 70^2 = 9800$ spins. This semi-classical phase space method is expected to be good in the collective regime, and has been shown to be accurate even for nearest neighbor interactions in two-dimensional systems \cite{Muleady_PRL_2023}. We provide benchmarks for the current model in the supplementary material \cite{supplemental}.


{\it Pair creation and Squeezing.---} %
In Fig. \ref{fig:fig2}(a) we show the generated excitations for infinite range ($\alpha=0$) and finite range ($\alpha=3$) interactions for the staggered z-field (dashed) and the Floquet engineered model (solid). The Floquet model for infinite range interactions creates the maximal number of possible excitations, $n_{ex} = 2N$, fully flipping the layer spin, in contrast to the staggered field case which saturates at a lower number due to the detuning of the pair creation process at finite number of excitations. The same advantage is observed for finite range interactions, with the Floquet model reaching a significantly higher number of generated pairs. %
The fully collective model ($\alpha=0$) here follows the TMS prediction (not shown), until corrections due to the finite spin-length become relevant. In contrast, the dipolar case ($\alpha=3$) at this layer separation $a_Z=2$ quickly deviates, which we understand to be due to non-collective spin dynamics as we will confirm below \cite{supplemental}. 
We next investigate the squeezing behavior under the same circumstances.  In Fig. \ref{fig:fig2}(b) we demonstrate the exponential decrease of the variance of the squeezed quadratures, $\mathrm{Var}[S^x_A + S^y_B]= \mathrm{Var}[S^y_A-S^x_B]\equiv \mathrm{Var}[\mathcal{O^-}]$, for different interaction ranges. %
We see that the spin dynamics follows the two-mode squeezing prediction $\mathrm{Var}[\mathcal{O}^{-}] =N/2 \, e^{- N V_{\mathrm{avg}}t/\hbar}$ up till a saturation point of minimal variance. Interestingly, the non-collective spin dynamics of the finite range model observed in the increased pair number does not seem to affect the variance, suggesting that non-collective spin excitations, affecting the spin polarisaiton, but not the variance, occur on top of the collective squeezing dynamics. While all models show identical short-time evolution, the saturation point depends on the model and interaction exponent $\alpha$. %
The Floquet model achieves a significantly lower variance, thus, a larger amount of squeezing and sensitivity for all interaction ranges. This is most notable for the infinite range case, where we observe a two orders of magnitude improvement, but holds true for all interaction ranges $\alpha$. The reduced improvement for finite range interactions is again explained by non-collective dynamics as we show next.

%
%

{\it Achieving the infinite range limit.---} %
\begin{figure}[t!]
\includegraphics[width=\columnwidth]{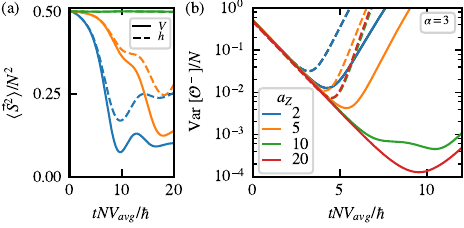}
\caption{Collective two-mode squeezing with finite range interactions. (a)  Length of collective layer spins $\langle \vec{S_A}^2 + \vec{S_B}^2\rangle$ with $\hat{S}^{\mu}_{A (B)} = \sum_{\boldsymbol{i} \in A (B)} \hat{s}_{\boldsymbol{i}}^{\mu}$ (b) Time evolution of variance of the squeezed quadrature. %
 In both panels for the Floquet-engineered model (solid lines, $V$), and for the staggered field model (dashed, $h$), for a range of layer spacings $a_Z$ with fixed powerlaw exponent $\alpha = 3$.
\label{fig:fig3}}
\end{figure} 
Despite the significant improvement of the achievable squeezing using the Floquet protocol, we observe that the finite range interaction cases do not saturate to the infinite-range model. This is due to the strong spatial inhomogeneity of the interlayer interaction at short interlayer separations $a_Z=2$, and the resulting non-collective excitations. We therefore explore the tunability of the layer spacing $a_Z$ to overcome this limitation in Fig. \ref{fig:fig3} in the case of dipolar interactions. %
We first study the collectiveness of the dynamics in terms of the length of the layer spins, %
$\langle \vec{S}_{\mathrm{A}}^2 + {\vec{S}_{\mathrm{B}}}^{2}\rangle$,
in Fig. \ref{fig:fig3}(a), demonstrating that as the layer-spacing increases the dynamics changes from a regime in which the collective layer spin rapidly dephases at short layer-distances, to fully collective behaviour at larger spacings where the layer spins stays fully collective with $\vec{S_\eta}^2=N^2/4$. We note that as expected longer-ranged interactions (smaller $\alpha$) achieve the collective regime for shorter layer distances, we provide data for $\alpha=1,2,3$ in the supplemental information \cite{supplemental}. %
As a direct consequence in Fig. \ref{fig:fig3}(b), which shows the time evolution of the variance of the squeezed quadrature, we observe that the achievable squeezing increases with layer distance. We note that the variance is actually more sensitive than the spin-length, showing increases of the minimal squeezing in a regime where the spin length appears fully collective already. Most importantly, the achievable squeezing in the dipolar ($\alpha=3$) saturates to the infinite range model (c.f. Figs~\ref{fig:fig2}(b) and \ref{fig:fig3}(b)) at sufficiently large layer spacings. Thus, finite range interactions are able to achieve the same amount of squeezing as infinite range, fully collective interactions.

{\it Heisenberg scaling of sensitivity.---} %
\begin{figure}
\includegraphics[width=\columnwidth]{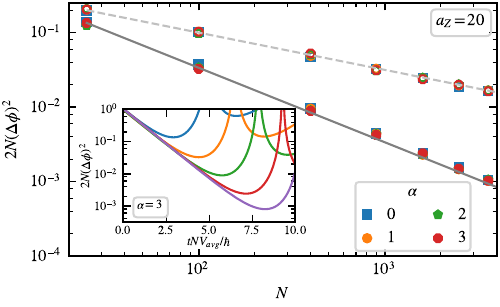}
\caption{Heisenberg scaling of sensitivity. Main panel: system size scaling of scaled sensitivity. Different symbols indicate exponent of interactions $\alpha$, filled symbols for Floquet-engineered model ($V$), open symbols for staggered field ($h$). Solid lines guide to the eye $1/\sqrt{N}$-scaling (light-gray) and Heisenberg $1/N$-scaling (dark grey). Inset: time-dependence of scaled sensitivity for different system sizes $L=5,10,20,40,70$ (top to bottom) at fixed $\alpha$ and $a_Z$ for the Floquet model. 
\label{fig:fig4}}
\end{figure} 
The squeezed variance of the generated entangled state directly results in an improved sensitivity of measuring a phase of rotation $\phi$ around $\hat{S}^z_A-\hat{S}^z_B$ in Ramsey protocols \cite{Sundar2023} as
\begin{equation}
(\Delta \phi)^2 =\frac{(\Delta \mathcal{O})^2}{ (\langle \hat{S}_A^z - \hat{S}_B^z \rangle)^2}
\end{equation}
and a corresponding enhancement over using $2N$ unentangled particles $2 N (\Delta \phi)^2$. As the state gets squeezed the variance decreases exponentially, however, at the same time the polarised spin component is reduced decreasing the sensitivity. Using the two-mode squeezing prediction for the variance and number of excitations the optimal scaled sensitivity is $27/(8 N)$ showing Heisenberg limited scaling \cite{supplemental}.

The inset of Fig.~\ref{fig:fig4} shows the time evolution of the scaled sensitivity for a range of different system sizes, demonstrating super-linear sensitivity gains beyond the standard quantum limit. In the main panel of Fig.~\ref{fig:fig4} we analyze the scaling of the optimal sensitivity with system size for both the staggered z-field (open symbols) and the Floquet-engineered model (filled symbols) for different power-law exponents $\alpha$. Notably, for large layer-spacings all power-law exponents collapse onto the same scaling. However, we observe two distinct scaling behaviours: the staggered field model shows $1/\sqrt{N}$ scaling, whereas the Floquet-engineered model achieves the optimal Heisenberg limit of $1/N$ scaling. Thus, using Floquet-engineered interactions provides a significant improvement in sensitivity for all system sizes and achieves the best possible scaling. %

{\it Outlook.---}
Our work demonstrates the scalable and robust generation of entanglement in the form of two-mode squeezed states separated in bilayers of powerlaw interacting quantum spin models. 

This extends the feasibility of two-mode squeezing to generic power law models, making it accessible in a  significantly larger number of experimental platforms. In particular, we show that finite range interactions ($\alpha=1,2,3$) can achieve the same amount of entanglement and squeezing as infinite range interactions ($\alpha=0$). 

We further develop a Floquet protocol utilizing spatio-temporal control to engineer the spin-spin interactions. This has a number of immediate benefits, it extends the applicability of our results to models with Ising interactions in systems, which may not naturally realize Heisenberg interactions. In addition, the Floquet engineered model achieves the optimal Heisenberg scaling of the sensitivity, providing potentially orders of magnitude improvements. Finally, it also allows to implement time-reversal by reversing the inter-layer spin-exchange interactions, which may be used for time-reversal based metrological protocols.

This establishes spatio-temporally engineered interactions adapted to the initial state and the desired dynamics as a viable pathway to unlocking significant quantum advantage beyond that present in naturally occuring interactions. It highlights the great potential in making full use of the control inherent in state-of-the-art current experimental platforms realizing fully controllable quantum spin systems for entanglement generation and quantum sensing.

 \begin{acknowledgments}
\noindent{\textit{Acknowledgements:}
The calculations were performed in the PETE system of the High Performance Computing Center at Oklahoma State University, NSF Grant No. OAC-1531128.
} 

\end{acknowledgments}



\bibliography{TMS_new}{}

\cleardoublepage
\appendix

\setcounter{equation}{0}
\setcounter{figure}{0}
\setcounter{table}{0}
\makeatletter
\renewcommand{\thefigure}{S\arabic{figure}}

\section{Supplementary Information}
The supplementary information contains additional details on the comparison of the dTWA simulations to exact numerics for finite and infinite-range ($\alpha=0$) interactions, convergence of the Floquet dynamics to the dynamics under the effective Hamiltonian, extended results on the behaviour of the collective layer spins for all interaction ranges, a comparison of the two-mode squeezing predictions with the spin dynamics for squeezed and anti-squeezed quadratures and for the sensitivity, and the dynamics for finite lattice filling fractions.

\subsection{Comparison of dTWA to exact results}
%
\begin{figure}
\includegraphics[width=\columnwidth]{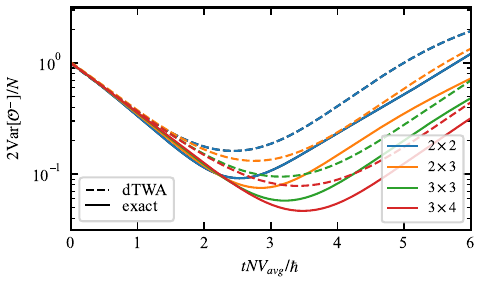}
\includegraphics[width=\columnwidth]{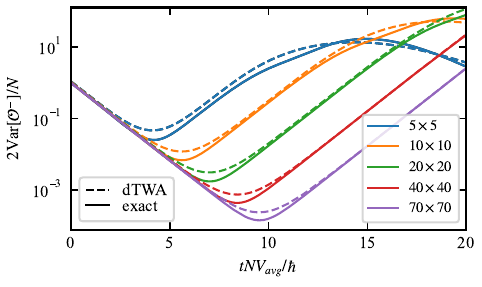}
\caption{Comparison with exact results. Variance $\mathrm{Var}[\mathcal{O}^{-}]$ comparing the dTWA results (dashed) to exact numerical results (solid) for the Floquet engineered model. Top: different layer sizes (see legend) for finite-range interactions $\alpha=3$. bottom: different layer sizes (see legend) for infinite range interactions $\alpha=0$. Both for $a_Z=2$. 
\label{SI_fig:fig1}}
\end{figure} 
In this section, we benchmark the results obtained using the dTWA method against exact diagonalization.

At finite $\alpha$ we are limited to small system sizes due to the rapidly growing Hilbert space of the problem. We compare the variance of the squeezed quadratues obtained from dTWA (dashed) to exact dynamics (solid) for a range of small bilayers in the top panel of Fig.~\ref{SI_fig:fig1}. We generally observe decent agreement, with very good agreement in the initial time-evolution, good agreement with the time-point of minimal variance, but observe that dTWA consistently seems to saturate at a too large minimal variance.

We additionally consider the model with infinite-range ($\alpha=0$) interaction, which can be solved using exact diagonalization on large systems. The dynamics starting from fully polarised layer spins remains in the Hilbert space of the fully collective states $\lvert S,m_A;S,m_B\rangle$, where $m_i=-S,\cdots,S$ with the additional constraint $m_A+m_B=0$, i.e. the relevant Hilbert space consists of only $2S+1$ total states, facilitating solutions for very large systems. In the bottom panel of Fig. \ref{SI_fig:fig1} we see that dTWA captures the evolution of the squeezed variance for large collective systems very well, with the main difference again being a slightly higher minimal variance. Importantly, this discrepancy seems to be a constant ratio of order 1, thus, not affecting the scaling predicted from dTWA simulations.

Thus, dTWA provides a good approximation to the time dynamics, and is likely to underestimate the achievable squeezing and sensitivity. The generally good agreement is consistent with the expectation that in the collective regime, the dTWA method captures the dynamics reliably.


\subsection{Comparison of effective hamiltonian and Floquet dynamics}
\begin{figure}
\includegraphics[width=\columnwidth]{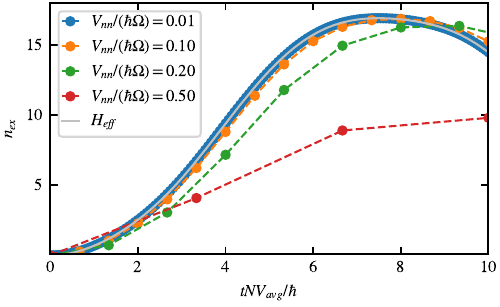}
\includegraphics[width=\columnwidth]{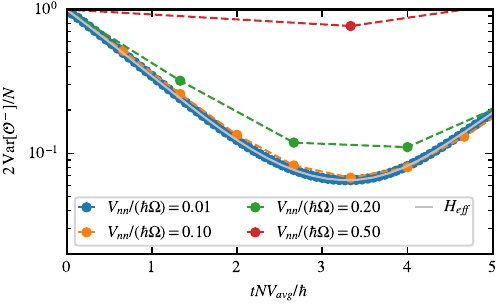}
\caption{Comparison of exact dynamics under the effective Hamiltonian and the time-dependent Hamiltonian. Time evolution of the generated entangled particles $N_{\mathrm{ex}}$ (top panel) and variance of the squeezed quadrature (bottom panel). Markers are the Floquet dynamics at the end of a full Floquet period for a range of periods $T$ characterised by the corresponding frequency $\Omega=2\pi/T$, the gray solid line is the effective Hamiltonian.  System parameters are $L=3$, $a_Z=1$, $\alpha=3$.
\label{SI_fig:fig_Floquet}}
\end{figure} 
In this section we analyize the convergence of the pulse-engineered dynamics to the dynamics generated by the effective Hamiltonian, Eq.~\ref{eq:Vz}.

To this end we simulate the exact dynamics under the time-dependent Hamiltonian 

\begin{equation}
H(t) =\begin{cases} H_{\mathrm{intra}}^{XX} + H_{\mathrm{inter}}^{XX}  \, \quad t \in [0,T/6) \\
                    H_{\mathrm{intra}}^{YY} + H_{\mathrm{inter}}^{YY}  \, \quad t \in [T/6,2T/6) \\
                    H_{\mathrm{intra}}^{ZZ} + H_{\mathrm{inter}}^{ZZ}  \, \quad t \in [2T/6,3T/6) \\
                    H_{\mathrm{intra}}^{ZZ} - H_{\mathrm{inter}}^{ZZ}  \, \quad t \in [3T/6,4T/6) \\
                    H_{\mathrm{intra}}^{YY} + H_{\mathrm{inter}}^{YY}  \, \quad t \in [4T/6,5T/6) \\
                    H_{\mathrm{intra}}^{XX} + H_{\mathrm{inter}}^{XX}  \, \quad t \in [5T/6,T) \\

      \end{cases}
\end{equation}
where $H^{\alpha \alpha}$ corresponds to the interaction terms $s_i^{\alpha} s_j^{\alpha}$ with $i,j$ in the same (different) layers for intra(inter)-layer interactions. \\

This is readily recognized as a step-drive Floquet Hamiltonian with period $T$ and associated frequency $\Omega = \frac{2\pi}{T}$. The lowest order effective Hamiltonian that describes the dynamics at multiples of the Floquet time-period $T$ is the desired target Hamiltonian 
\begin{equation}
\begin{split}
H_{\mathrm{eff}} = \frac{1}{T} \int_0^T H(t) = 1/3 \left[ (H_{\mathrm{intra}}^{XX}+H_{\mathrm{intra}}^{YY}+H_{\mathrm{intra}}^{ZZ})  \right. \\
\left. +  ( H_{\mathrm{inter}}^{XX}+H_{\mathrm{inter}}^{YY})\right]
\end{split}
\end{equation}
The effective Hamiltonian is expected to provide a good description of the dynamics in the limit of the Floquet period being small compared to all interactions present in the Hamiltonian. We can ensure this condition by comparing the Floquet frequency to the largest interactions, which are the nearest neighbour interactions $V_{nn}$, resulting in the condition $V_{nn}/(\hbar \Omega) \ll 1$.

We exactly simulate the dynamics under the time-dependent Hamiltonian and this effective Hamiltonian for small systems and compare the results for the number of generated excitations and the variance of the squeezed quadrature in Fig.~\ref{SI_fig:fig_Floquet}. The markers correspond to the Floquet results after multiples of the Floquet period, and the continuous line shows dynamics under the effective Hamiltonian. We observe that the Floquet dynamics indeed converges towards the dynamics of the effective Hamiltonian with increasing Floquet frequency $\Omega$, and observe good agreement around $V_{nn}/(\hbar \Omega) = 0.1$ both for the generated excitations and the squeezing achieved.
In terms of the Floquet period this corresponds to $T\approx 0.6 V_{nn}/\hbar$, i.e. about 2 pulse-sequences per nearest-neighbour interaction time. This is readily achievable in current experiments, e.g. \cite{miller2024twoaxis}.

This demonstrates that the proposed Floquet protocol indeed allows the simulation of the desired target Hamiltonian.

\subsection{Collectiveness}
\begin{figure}
\includegraphics[width=\columnwidth]{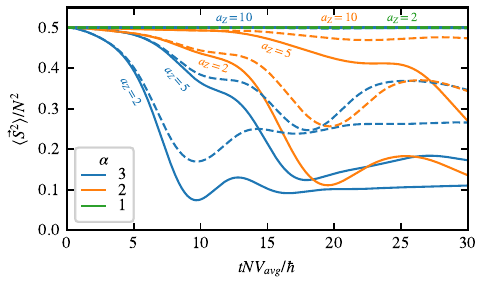}
\caption{Collectiveness of dynamics. Sum of length of collective layer spins $\langle \vec{S_A}^2 + \vec{S_B}^2\rangle$ with $\hat{S}^{\mu}_{A (B)} = \sum_{\boldsymbol{i} \in A (B)} \hat{s}_{\boldsymbol{i}}^{\mu}$ as a function of evolution time for a range of layer spacings $a_Z$ and interaction powerlaw exponents $\alpha$. System size $L=70$.
\label{SI_fig:fig2}}
\end{figure} 
\begin{figure}
\includegraphics[width=\columnwidth]{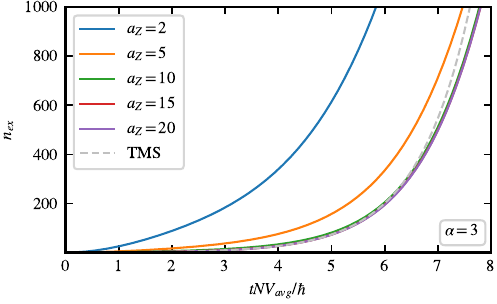}
\caption{Number of created excitations $n_{ex} =S_A^z - S_B^z+N$ for $\alpha=3$ for different layer spacings $a_Z$ (solid lines) compared to the TMS prediction (dashed), $n_{ex}=2\sinh^2(NV_{avg}t/(2 \hbar)$. System size $L=70$ corresponding to $N=4900$.
\label{SI_fig:Npair_collectiveness}}
\end{figure} 
We have shown that the tunability of interlayer distance, $a_Z$, plays an important role in achieving maximal squeezing. This is due to the direct influence $a_Z$ has on the inhomogeneity of the interlayer interactions. A larger layer separation gives rise to a more homogeneous interlayer interaction, reducing interaction terms that would create excitations outside the collective manifold of maximal layer spin length. %

Thus, at larger layer separations we expect the layer spin length, $\vec{S}_{\eta}$, to be protected, enabling maximal achievable squeezing. In Fig. \ref{SI_fig:fig2} we verify this by showing the transition to collective behavior with increasing $a_Z$ for finite range ($\alpha = 1, 2, 3$) interactions. %
As expected, for short-range interaction ($\alpha=3$), with a high degree of spatial inhomogeneity of the interactions, the initially prepared state quickly loses its collective behavior and the layer spin dephases for small layer distances $a_Z$. In contrast, as the layer distance increases, the system stays fully collective with the layer spin retaining maximal spin length $N^2/4$ in each layer. For longer-ranged interactions ($\alpha=1$), this transition occurs at smaller layer distances, and the dynamics remains collective even at a small layer separation of $a_Z=2$. 

This transition is reflected in the dynamics of the squeezed variances and their saturation towards the infinite range case, which is always fully collective, seen in Fig. \ref{fig:fig3}. We note that while it might appear that the staggered z-field model has a slight advantage here, in that it shows a larger spin length after dephasing, this can be mainly explained by the fact that the detuning effectively restricts the accessible Hilbert space by limiting the number of excitations. This does not translate in an advantage for the achievable squeezing, which occurs only in the fully collective regime here.

In addition to the effects on the variance, we also observe this transition in the dynamics of the created spin excitations, $n_{ex} =S_A^z - S_B^z+N$, which were seen to not follow the TMS prediction for finite range interactions at small layer spacings in Fig.~\ref{fig:fig2}(a). We confirm that the pair creation dynamics indeed converges towards the TMS prediction in Fig.~\ref{SI_fig:Npair_collectiveness} with increasing layer distance/collectiveness. Deviations appear when the number of excitations becomnes comparable to the total particle number, i.e. for $N_{\mathrm{ex}} \sim 0.1 N$

\subsection{Squeezed and anti-squeezed quadrature}
\begin{figure}
\includegraphics[width=\columnwidth]{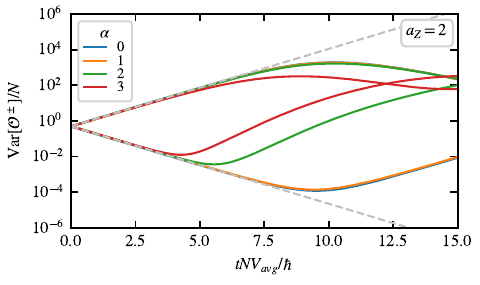}
\caption{Squeezed and anti-squeezed quadrature. dTWA results (solid), and two-mode squeezing prediction (dashed), $\mathrm{Var}[\mathcal{O}^{\pm}] =N/2 e^{\pm N V_{\mathrm{avg}}t/\hbar}$, for $L=70$, $a_Z=2$ for a range of powerlaw exponents $\alpha$ as indicated in the legend.
\label{SI_fig:fig3}}
\end{figure} 

We compare the exponential decrease (increase) of variance for the squeezed (anti-squeezed) quadrature, $\mathrm{Var}[\mathcal{O}^{\pm}] =N/2 e^{\pm N V_{\mathrm{avg}}t/\hbar}$, corresponding to exponential creation in time of entanglement, predicted by the two-mode squeezing Hamiltonian in this section, to the results of the dTWA simulation. 

In Fig. \ref{SI_fig:fig3} we plot both the squeezed and anti-squeezed variances for different powerlaw exponents. At short time, the variance agrees with the TMS prediction perfectly for all $\alpha$ but starts deviating as time evolves, with the shortest range interactions showing deviations the earliest. This is explained by the fact that for shorter-range interaction ($\alpha = 2,3$), the initial collective layer spins lose their collectiveness more quickly (see Fig. \ref{SI_fig:fig2}) invalidating the TMS prediction. For long and infinite range interactions, the spins in the layers remain collective and the variance agrees with TMS prediction up to close to its minimal value, where the spin dynamics slows down, before reversing. In the fully collective regime, the slow down of the spin dynamics compared to the prediction within first order Holstein-Primakoff bosons is explained by the fact that as excitations are created the effective interaction constant appearing in the two-mode squeezing hamiltonian is decreased by higher order corrections from $S/2$ to $S/2 - n_{ex}/2$. Ultimately, the collective spin model has a finite-dimensional Hilbert space of size $2S+1$ compared to the infinite dimensional Hilbert space of bosons, which will naturally limit the pair creation process. In fact, for the Floquet engineered Hamiltonian, taking into account the symmetry under $\hat{S}^z \rightarrow - \hat{S}^z$, which exchanges layers A and B, the dynamics will reverse after crossing $S^Z_{A,B} = 0$ and will reach a fully uncorrelated state when both layers have been fully flipped compared to the initial state. This explains the oscillatory and symmetric time dynamics of the variances in the fully collective regime. For finite range interactions in the non-collective regime, the dynamics deviates from the two-mode squeezing prediction earlier, with the squeezed and anti-squeezed quadratures not showing perfectly symmetric behaviour, and the dynamics does not fully flip the layer spins.

\subsection{Sensitivity}
\begin{figure}
\includegraphics[width=\columnwidth]{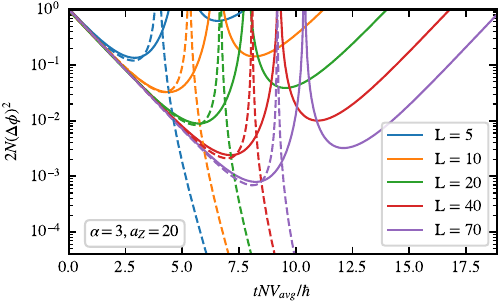}
\caption{Scaled sensitiivity as defined in Eq.~\ref{eq:gain} for $\alpha=3$, $a_Z=20$ for a range of system sizes. dTWA results (solid-lines) compared to the two-mode squeezing prediction, Eq.~\ref{eq:TMS_gain}, (dashed lines). 
\label{fig:Metrological Gain}}
\end{figure} 
In this section we consider the sensitivity of estimating a rotational phase $\phi$ using the two-mode squeezed spin state in a Ramsey type experiment. Following the protocol described in \cite{Sundar2023} this is given by 
\begin{equation}
(\Delta \phi)^2 = \frac{(\Delta \mathcal{O})^2}{ (\langle S_A^z - S_B^z \rangle)^2}
\end{equation}
and the corresponding enhancement over using an unentangled state of $2 N$ atoms is consequently given as
\begin{equation}
  2 N (\Delta \phi)^2 =\frac{2N \, (\Delta \mathcal{O})^2}{ (\langle S_A^z - S_B^z \rangle)^2}
\label{eq:gain}
\end{equation}
Using the two-mode squeezing predictions, $\mathrm{Var}[\mathcal{O}^{\pm}] =N/2 \, e^{\pm N V_{\mathrm{avg}}t/\hbar}$, and,  $n_{ex} = S_A^z - S_B^z+N= 2\sinh^2(N V_{\mathrm{avg}} t/(2\hbar)) $, this becomes
\begin{equation}
2N (\Delta \phi)^2 = \frac{N^2 \, e^{-N V_{\mathrm{avg}}t/\hbar} }{(N - 2\sinh^2(N V_{\mathrm{avg}} t/(2\hbar)))^2}
\label{eq:TMS_gain}
\end{equation}
Clearly the sensitivity, and the improvement over the unentangled state, is reduced by the creation of pairs, shrinking the polarised spin component. 

Importantly, this does not change the resulting scaling of the optimal sensitivity with $N$. Intuitively, if the minimum is attained after a fraction $ r N$ entangled particles are created, the denominator only results in a prefactor of $1/(1-r)^2$. Explicitly optimising Eq.~\ref{eq:TMS_gain} over $t$ results in an asymptotic scaling $t_{opt} \sim \mathrm{log}(2 N/3)$, $n_{ex,opt} \sim N/3$, and an optimal scaled sensitivity $\sim 27/(8 N)$, still showing Heisenberg scaling.

We compare in Fig.~\ref{fig:Metrological Gain} the results of the dTWA simulations for the metrological gain, Eq.~\ref{eq:gain}, and the two-mode squeezing prediction, Eq.~\ref{eq:TMS_gain}. We observe that the spin dynamcis indeed closely follows the TMS prediction, including the maximally achievable sensitivity. 

\subsection{Robustness to Disorder}
\begin{figure}
\includegraphics[width=\columnwidth]{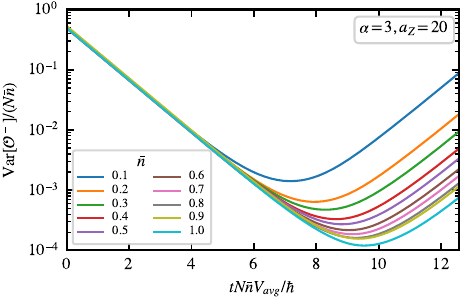}
\caption{Robustness to disorder/non-unit filling. Time evolution of the squeezed variance for a range of filling fractions $\bar{n}$ at fixed $\alpha=3$, $a_Z=20$, $L=70$ for the Floquet engineered model.
\label{SI_fig:fig4}}
\end{figure} 
In this section, we demonstrate the robustness of our predictions to fractional filling of the lattice layers. We simulate the dynamics of the spin model assuming a random configuration of $\bar{n} N$ sites per layer to be occupied, with the remaining sites being empty. This random static disorder in the positions of occupied sites then for spatially dependent interactions directly results in disordered interactions. %
%
%
We average the simulations results over 20 realizations of the site occupations, which ensures convergence of the variances shown in the following.

In Fig. \ref{SI_fig:fig4} we plot the time evolution of the squeezed variance for a range of filling fractions starting from unit filling down to only $10\%$ for dipolar $\alpha=3$ interactions. We note that since we expect the main effect of finite filling fraction to be the disordered interactions, using the most short-range case results in the most stringent test of the robustness. We observe that the results collapse when the time axis is scaled by $\bar{n} N$. This indicates that the dynamics remains fully collective, and is still given by the average interaction each spins feels, which is trivially reduced by the number of available interaction partners in a non-unit filled lattice. 



\end{document}